\documentclass[twocolumns]{aa} 

\usepackage{graphicx}
\usepackage{txfonts}
\usepackage{hyperref}
\usepackage{multirow}
\usepackage{comment}

\usepackage{natbib}
\begin{document}

   \title{JWST uncovers helium and water abundance variations in the bulge globular cluster NGC~6440}

     \author{Mario~Cadelano \inst{1,2}
          \and
          Cristina~Pallanca \inst{1,2}
          \and
          Emanuele~Dalessandro \inst{2}
          \and
          Maurizio~Salaris\inst{3,4}
          \and
          Alessio~Mucciarelli \inst{1,2}
          \and
         Silvia~Leanza \inst{1,2}
         \and
         Francesco~R.~Ferraro \inst{1,2}
         \and
         Barbara~Lanzoni \inst{1,2}
         \and
         Rosie~H.~Chen \inst{5}
         \and
         Paulo~C.~C.~Freire \inst{5}
         \and 
         Craig Heinke \inst{6}
         \and
         Scott M. Ransom \inst{7}
          }

   \institute{Dipartimento di Fisica e Astronomia “Augusto Righi'', Universit\`a degli Studi di Bologna, Via Gobetti 93/2, 40129 Bologna, Italy \\
         \and
             INAF-Osservatorio di Astrofisica e Scienze dello Spazio di Bologna, Via Gobetti 93/3 I-40129 Bologna, Italy \\
             \and
             Astrophysics Research Institute, Liverpool John Moores University, 146 Brownlow Hill, Liverpool L3 5RF,UK \\
             \and
             INAF - Osservatorio Astronomico d'Abruzzo, via M. Maggini, 64100, Teramo, Italy \\
             \and
             Max-Planck-Institut für Radioastronomie MPIfR, Auf dem Hügel 69, D-53121 Bonn, Germany
             \and
             Department of Physics, University of Alberta, Edmonton, AB T6G 2G7, Canada
             \and
             NRAO, 520 Edgemont Road, Charlottesville, VA 22903, USA
             }

   \date{Received September 15, 1996; accepted March 16, 1997}

 
  \abstract{
 We used ultra-deep observations obtained with the NIRCam aboard the James Webb Space Telescope to explore the stellar population of NGC~6440: a typical massive, obscured and contaminated globular cluster formed and orbiting within the Galactic bulge. 
Leveraging the exceptional capabilities of this camera, we sampled the cluster down to $\sim5$ magnitudes below the main-sequence turn-off in the
($m_{F115W},m_{F115W}-m_{F200W}$) colour-magnitude diagram.
After carefully accounting for differential extinction and contamination by field interlopers, we find that the main sequence splits 
into two branches both above and below the characteristic knee.
By comparing the morphology of the colour-magnitude diagram with a suitable set of isochrones, we argue that the upper main-sequence bi-modality is likely due to the presence of a He-enriched stellar population with a helium spread of $\Delta Y=0.04$. 
The lower main-sequence bi-modality can be attributed to variations in the abundance of water (i.e., oxygen) with $\Delta [O/Fe]\sim-0.4$. 
This is the first 
evidence of both helium and oxygen abundance variations in a globular cluster purely based on JWST observations. These results open the window for future in-depth investigations of the multiple population phenomenon in clusters located in the Galactic bulge,  which were previously unfeasible with near-UV observations, due to prohibitive reddening and crowding conditions.}

    
    \titlerunning{Multiple populations in NGC~6440}
   \maketitle

%

\section{Introduction}\label{sec:intro}

The existence of multiple stellar populations (MPs) characterized by variations in the abundance of light-elements (such as He, C, N, O, Na, Mg, Al), while having the same iron-peak content, is a fundamental characteristic observed in virtually all massive and old ($M>10^4 M_{\odot}$, $t\gtrsim 1.5$ Gyr) globular clusters (GCs) within the Milky Way and other galaxies \citep[see, e.g.,][]{Bastian2018, gratton19,piotto15,mucciarelli08,larsen14,Dalessandro2016,Martocchia2018,sills19,cadelano2022}. Stars exhibiting light-element abundance ratios resembling those of the surrounding field (i.e., Na-poor/O-rich) are referred to as first-population (FP) stars, while those with Na-rich/O-poor abundances are known as second-population (SP) stars. 

MPs can be studied through photometry due to the effects that different light-element abundances induce on stellar effective temperatures, luminosities, and spectral energy distributions \citep[e.g.][]{salaris06,salaris19,Sbordone2011,cassisi20}. These result in splitting or spreads of the evolutionary sequences in the colour-magnitude diagram (CMD) when appropriate filter combinations are adopted \citep[e.g.][]{piotto2015,Milone2017,milone2019,onorato23}. 
In this respect, the Hubble Space Telescope (HST) with its high-resolution and broad-band capabilities has brought a revolution in the field, significantly enhancing our understanding of the chemical, structural, and kinematic properties of these sub-populations (e.g., \citealt{piotto15,Milone2017,Bellini2015,dalessandro18_m80,dalessandro2019}). With the recent launch of the James Webb Space Telescope (JWST), a new window of opportunity has opened to further investigate 
this phenomenon. In fact, as extensively discussed by \citet[][see also \citealt{nardiello22}]{salaris19}, 
JWST filters offer the potential to detect characteristic abundance variations associated with the MPs along both the RGB and the main-sequence. Specifically, the near-IR filters employed by the JWST cameras are particularly sensitive to effective temperature variations caused for example by He abundance variations, and to the abundance of water molecules of M-dwarfs \citep[see also][]{milone12,milone2019}, which correspond to variations in oxygen content, one of the main light elements involved in the MP phenomenon.

Several scenarios have been proposed over the years to explain the formation of MPs, however, their origin is still debated  \citep[see, e.g.,][]{Bastian2018,renzini22}. Exploring the MP phenomenon in different galactic environments may provide valuable insights into the many physical processes contributing to the observed chemical inhomogeneities. {In this respect, a systematic mapping of MPs in bulge GCs is still missing due to observational challenges. The combination of severe extinction, crowding and contamination by field interlopers have hindered previous attempts to explore MPs both through spectroscopy of large stellar samples, and through optical-UV photometry. However, bulge GCs constitute a large fraction of the in-situ formed cluster population \citep{massari19}, and their MP properties likely retain valuable insights on the environmental conditions in which they were formed.

This Letter presents the first photometric evidence of MPs in NGC~6440 obtained by using JWST/NIRcam observations. NGC~6440 is a typical high-mass and metal-rich GC formed and orbiting within the Galactic bulge. Extensive studies of this cluster have revealed a dynamically evolved system, located at 8.3 kpc from the Sun, with a high-mass of $2.7\times10^5 \, M_{\odot}$, affected by a severe degree of differential reddening ($\delta E(B-V)\sim0.5$ mag), which evidently distorts the observed evolutionary sequence in the CMDs 
\citep{cadelano17,pallanca19,pallanca21,leanza23,ferraro23}. The first evidence of MPs in NGC~6440 was found through spectroscopic analysis of seven giant stars by \citet{munoz17}. The authors observed intrinsic significant variations in the abundances of Na and Al, while no significant spread in the O abundance was detected. 

Here we demonstrate how the stand-alone exceptional photometric capabilities of the JWST have made possible the detection of MPs  in terms of both helium and water abundance variations in such a challenging  system representative of the bulge GC population. 
}

\section{Data-set and data reduction} \label{sec:data}
We used data obtained with the JWST Near Infrared Camera (NIRCam) as part of the GO 2204 program (PI: Freire) conducted during Cycle~1. The dataset includes 20 images acquired with the F115W filter {(Pivot wavelength $\lambda_p=1.154 \mu m$)}, with exposure time of 376 s each, and 20 images acquired with the F200W filter ($\lambda_p=1.990 \mu m$), with exposure time of 215 s each. To complement these observations, we also used the optical data obtained from the Wide Field Camera 3 (WFC3) onboard the HST using a combination of F606W and F814W filters \citep[see][]{pallanca19,pallanca21}. A detailed and comprehensive description of the specific tools and strategies adopted for the analysis of the observations will be described in detail in a forthcoming paper (Pallanca et al., in prep.). 
Here we briefly summarize the main steps. We used DAOPHOT II \citep{Stetson1987,Stetson1994} to model the point spread function independently for each chip and band and to perform a simultaneous fit to all the sources detected above an appropriate threshold in the JWST and HST observations. The final catalog includes homogenized magnitudes for all available filters. They were reported to the VEGAMAG photometric system by adopting appropriate zero-points and aperture corrections. 

We optimized the proper motion analysis originally presented in \citet{pallanca19} to incorporate the new JWST epoch. The calibrated magnitudes were corrected for the heavy degree of differential reddening affecting the field-of-view. This was estimated 
by using the approach described in \citet[][see also \citealt{cadelano20_n6256}]{pallanca19} taking advantage of the NIR-optical filter combination
enabled by the simultaneous use of JWST and HST. {Briefly, the estimation of differential reddening consisted in two steps. First, we modeled the cluster's mean ridge line using a set of bona-fide cluster members (see Figure~4 in \citealt{pallanca19}). Then, in the second step, we measured the color excess variation
$\delta E(B-V)$ for each star with respect to the cluster's average value. This measurement consists in determining the shift along the reddening vector (defined using the extinction coefficients by \citealt{cardelli89,O'Donnell1994}) required to minimize the CMD distance between the mean ridge line and a sample of up to 50 cluster members located within a  maximum distance of $5\arcsec$ from the investigated star. The observed field of view turned to be affected by color excess variations up to $\delta E(B-V)\sim0.6$ mag and the resulting differential reddening map is in excellent agreement with the one published by \citet{pallanca19} using HST data.}
We refer to  Pallanca et al., in prep. for a  comprehensive description of both the revised differential reddening correction and the proper motion analysis.



\section{Discovery of a dual bi-modal main-sequence} \label{sec:ms}
\begin{figure}[t] 
\centering
\includegraphics[scale=0.43]{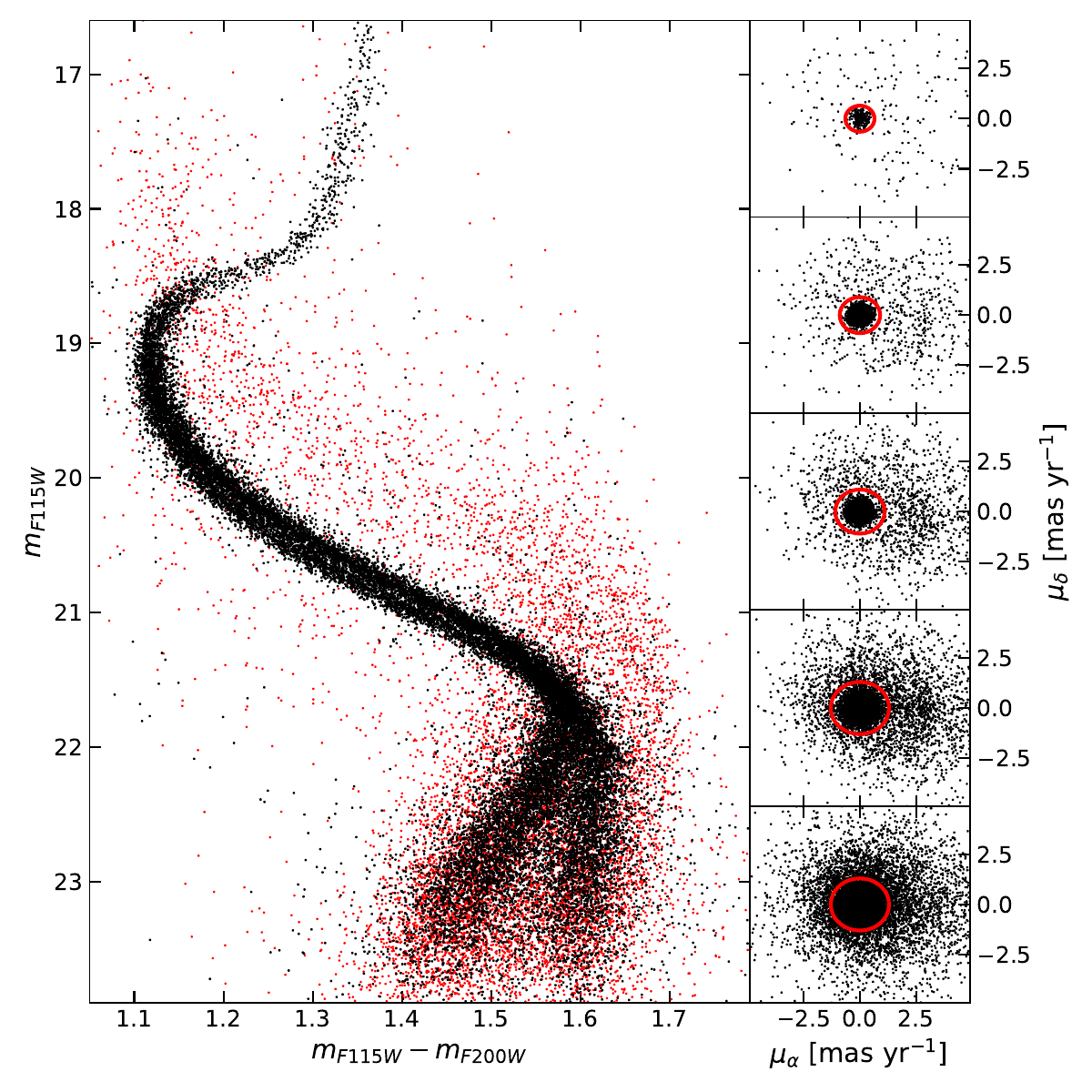}
\caption{{\it Left-hand panel:} CMD of NGC~6440 obtained with a combination of the F115W and F200W NIRCam@JWST filters. Black and red dots are high-quality photometry stars flagged as cluster member and field interlopers, respectively.  {\it Right-hand panels}: Vector-point-diagram (i.e. proper motion along RA vs proper motion along Dec) in different magnitude ranges: stars within and beyond the red circles are flagged as cluster members and field interlopers, respectively.}   \label{fig1}
\end{figure}

To perform a  detailed analysis of the CMD features of the cluster, we selected a sample of high-quality and bona-fide cluster member stars in the final photometric catalog. Initially, we excluded stars with photometric errors, chi-square values, and sharpness measurements that exceeded $3\sigma$ of the local magnitude values for each filter. 
Then, we eliminated stars with uncertain differential reddening value according to the residuals of the correction and those located within a distance smaller than the cluster's half-mass radius ($r_{hm}=50\arcsec$; \citealp{pallanca21}). The latter exclusion mitigated potential issues related to blending and saturation that typically arise in the densely populated inner regions of the cluster. Finally, we excluded stars having proper motions larger than three times the dispersion calculated in the corresponding magnitude bin (Figure~1). 
The $(m_{F115W}, m_{F115W}-m_{F200W})$ colour-magnitude diagram of the selected sample of stars is shown in Figure~\ref{fig1} along with the vector-point-diagrams used to remove field interlopers.

The CMD obtained from the JWST observations (see left-hand panel of Figure~\ref{fig1}) extends for more than 8 magnitudes and samples the main sequence down to approximately 5 magnitudes below the turn-off point. Notably, two prominent features are observed along the cluster main-sequence. A bi-modal structure in the magnitude range of $20 < m_{F115W} < 21.25$ is observed as two nearly parallel and almost equally populated sequences of stars. 
Such a bi-modality becomes less evident for magnitudes brighter than $m_{F115W}=20$ and it disappears around $m_{F115W}=21.25$. 
Below the so-called main-sequence knee (i.e., at $m_{F115W} > 22$), a second 
bi-modal pattern becomes apparent
in the form of a bluer diagonal sequence and a redder nearly vertical branch. 
This double bi-modality feature, observed simultaneously in the same filter combination at magnitudes brighter and fainter than the main-sequence knee, provide compelling evidence for the presence of MPs within this cluster.  

\begin{figure*}[!h] 
\centering
\includegraphics[scale=0.5]{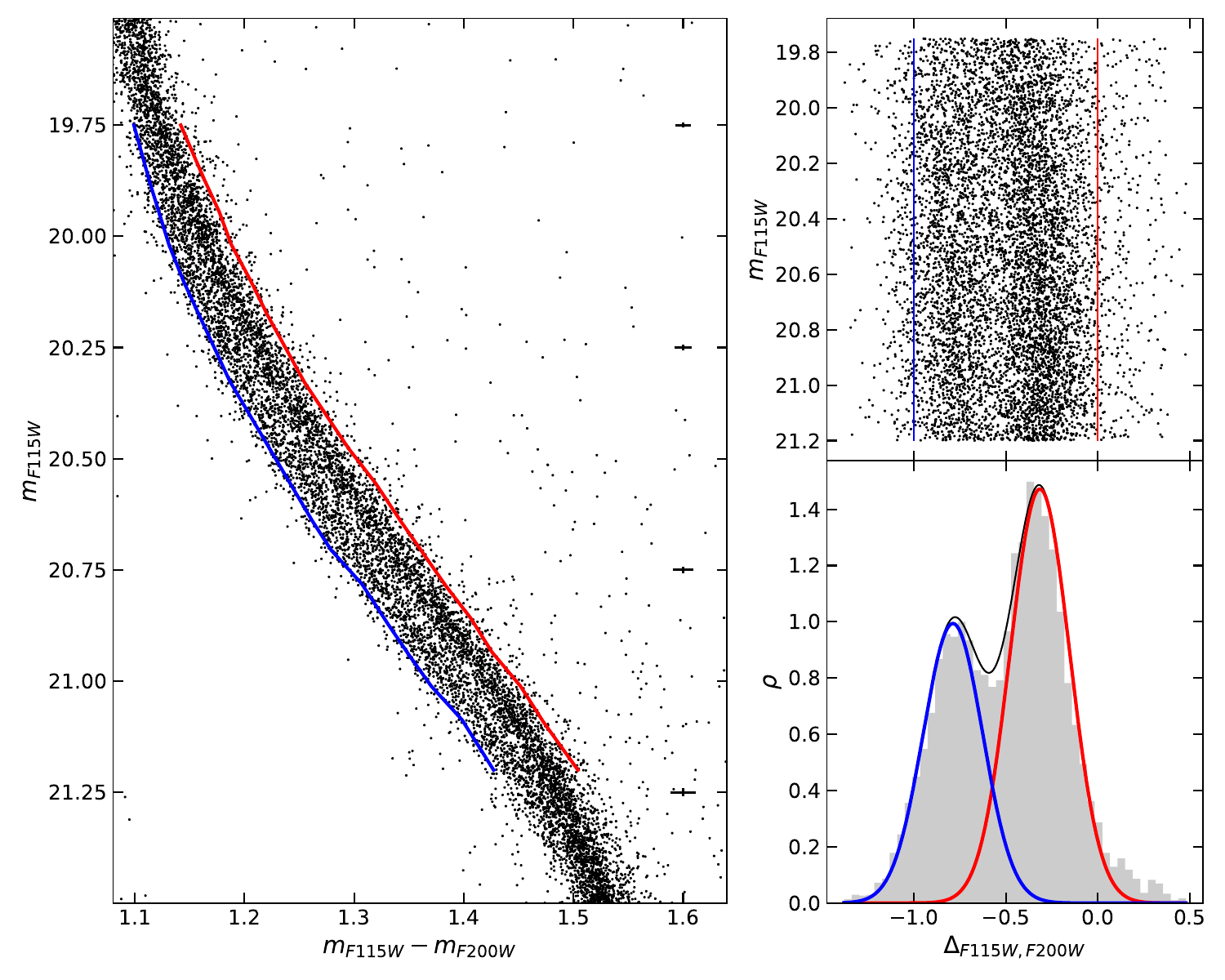}
\caption{{\it Left-hand panel:} JWST CMD of NGC~6440 zoomed on the upper main-sequence. The red and blue curves are the two fiducial lines adopted to verticalize the colour distribution. The error bars on the right side on the plot represent the average photometric uncertanties.  {\it Right-hand panel:} The top panel shows the verticalized colour distribution of main-sequence stars in the magnitude range $19.75<m_{F115W}<21.2$. The bottom panel shows instead the histogram of the same distribution. The blue and red curves are the best-fit Gaussian functions to the observed  distribution, while the black one is the sum of them.}  
\label{fig2}
\end{figure*}

To analyze the structure of the main-sequence in more detail, we verticalized the colour distribution of the stars in the magnitude range $19.7 < m_{F115W} < 21.2$ with respect to two fiducial lines {following the blue and red ends of the sequence (see Figure~\ref{fig2}) and calculated as the $5^{th}$ and $95^{th}$ percentile of the colour distribution in different magnitude bins} (see \citealt{dalessandro2018} for a similar implementation of the technique). The left-hand panel of Figure~\ref{fig2} displays a zoom on the CMD and the two fiducial lines. The resulting verticalized distribution and its corresponding histogram are presented in the right-hand panels. Notably, the verticalized distribution exhibits two distinct and clearly separated sequences, which are also confirmed by the two strikingly separated peaks in the histogram. 
We performed a Dip-test \citep{dip85}  and found out that the probability that this colour distribution is uni-modal is lower than 0.5\%.  To better characterize this bi-modal distribution, we applied a Gaussian Mixture model statistic with two components to fit the histogram of the verticalized colour distribution. The resulting best-fit model is represented by the red and blue curves in the corresponding figure. According to the best-fit analysis, the red sequence includes 4414 stars, while the blue main-sequence contains 2913 stars, corresponding to 60\% and 40\% of the total sample, respectively. 

\begin{figure}[h] 
\centering
\includegraphics[scale=0.43]{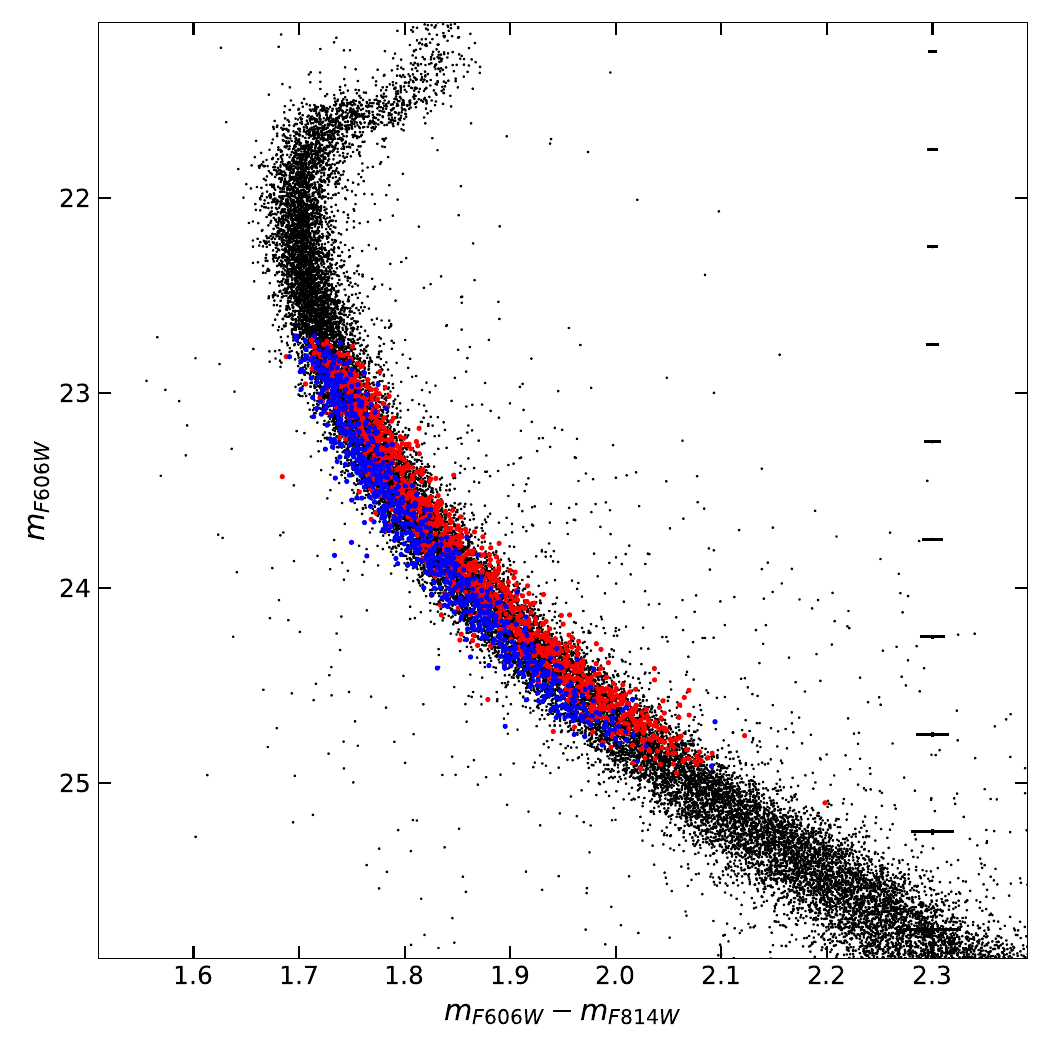}
\caption{Optical CMD of NGC~6440 obtained through the combination of the F606W and F814W filters of the HST/WFC3. Magnitudes are corrected for differential reddening and only bona-fide cluster members are plotted. The blue and red stars are those belonging to the blue and red main-sequence in the JWST data components selected through the Gaussian mixture model fit (see Figure~\ref{fig2}).}  
\label{fig3}
\end{figure}

We also investigated the positions of the red and blue sequence stars in the purely optical HST (F606W,F814W) CMD, which is a classical filter combination mostly sensitive to effective temperature variations. Interestingly, as depicted in Figure~\ref{fig3}, the colour separation between the two sequences is preserved also in this frame, although the separation is smaller and less clearly evident. 

\begin{figure*}[h] 
\centering
\includegraphics[scale=0.5]{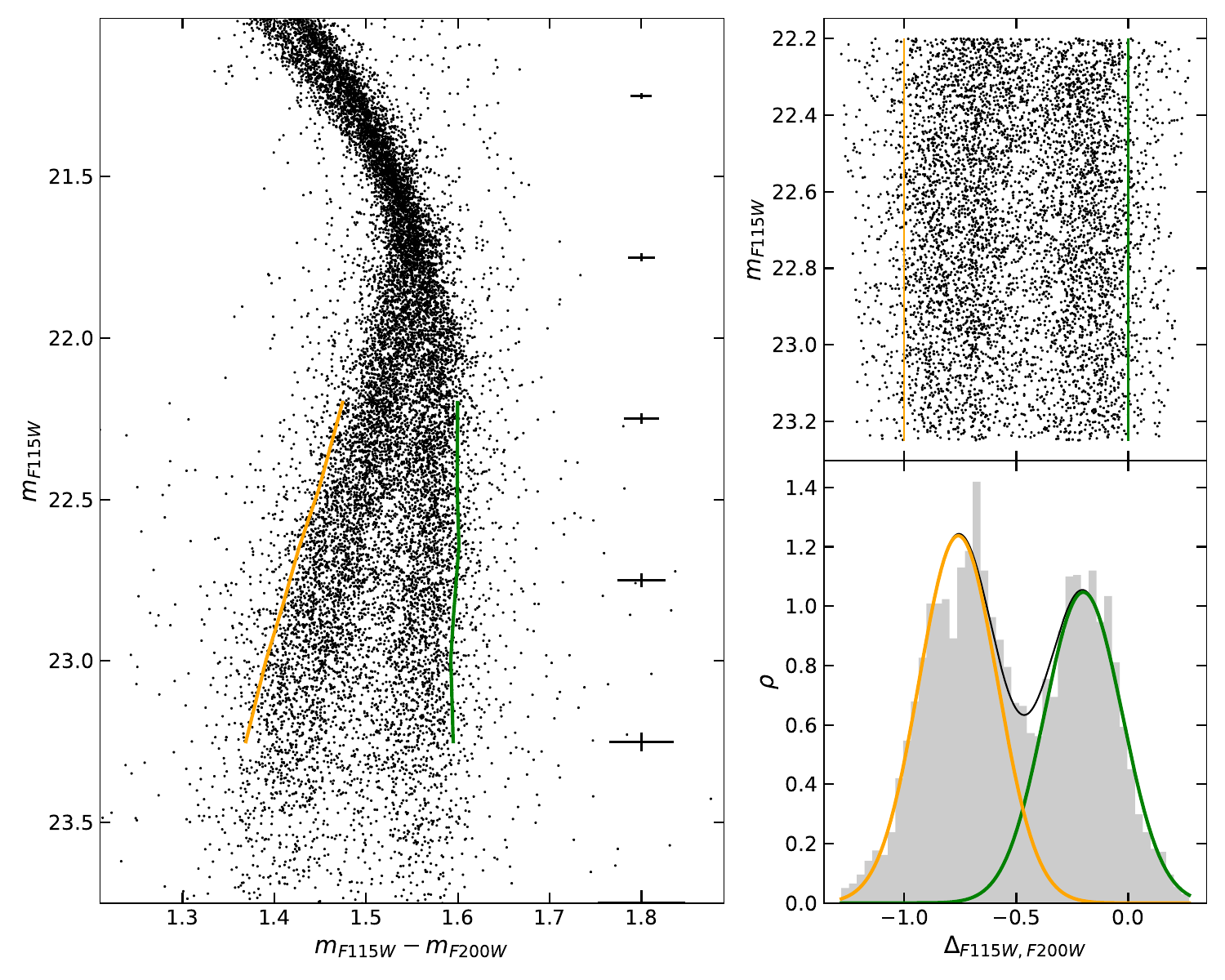}
\caption{{\it Left-hand panel:} JWST CMD of NGC~6440 zoomed on the lower main-sequence. The orange and green curves are the two fiducial lines adopted to verticalize the colour distribution. The error bars on the right side on the plot represent the average photometric uncertanties.  {\it Right-hand panel:} The top panel shows the verticalized colour distribution of main-sequence stars in the magnitude range $22.2<m_{F115W}<23.25$. The bottom panel shows instead the histogram of the same distribution. The orange and green curves are the best-fit Gaussian functions to the observed  distribution, while the black one is the sum of them.}  
\label{fig4}
\end{figure*}

To analyze the bi-modality below the main-sequence knee and specifically in the magnitude range $22.2<m_{F115W}<23.25$, we applied the same approach discussed above. The results are presented in Figure~\ref{fig4}. The verticalized distribution confirms the presence of at least two prominent and nearly equally populated structures: a blue and diagonal component and a red and almost vertical one. 
The Dip-test yielded a null probability, indicating that the colour distribution is not uni-modal. Fitting the distribution with two Gaussian components, we find that $3124$ stars belong to the blue sequence, while $2583$ stars belong to the red vertical sequence, corresponding approximately to 55\% and 45\% of the total sample. 
It is noteworthy, however, that the blue diagonal component seems to exhibit further substructure. Indeed, two slightly separated structures are discernible both in the CMD, and in the verticalized colour distribution. In the latter, the separation occurs at $\Delta_{F115W, F200W}=-0.78$ and is robust against variations in the histogram binning. Although the Dip-test returns a $\sim 80\%$ probability that the colour distribution is uni-modal, future observations will hopefully shed light in the possibility that an intermediate sub-population is also present in this cluster. 

Finally, due to the severe incompleteness and low signal-to-noise ratio of the common stars in the HST frame (particularly in the F606W filter), a direct comparison in the optical frame is not feasible.

\section{Comparison with models} \label{sec:iso}

The most plausible explanation of the presence of the two bi-modal distributions observed in this cluster is the co-existence of two stellar populations with different light-element abundances, the SP being composed not only of N-rich/O-poor stars but also significantly enriched in helium. Above the main-sequence knee, variations in nitrogen or oxygen abundance would have a minimal impact on the colours. Therefore, the observed split is more likely a result of effective temperature variations, which are commonly associated with helium and/or iron abundance variations. 
Assuming that NGC~6440 has no intrinsic metallicity spread \citep{origlia08,munoz17}, the most plausible interpretation is that the observed split in the upper main-sequence is indeed caused by a significant difference in the He mass fraction ($Y$) between the two sub-populations.

On the other hand,  the bi-modal split observed at magnitudes fainter than the knee, is  probably linked to variations in the abundance of water (i.e., oxygen) on the surfaces of the stars. In fact, the F200W filter is particularly sensitive to oxygen abundance variations since it covers a wavelength range where molecular absorptions due to the water molecule fall \citep{salaris19}.

\begin{figure*}[h] 
\centering
\includegraphics[scale=0.5]{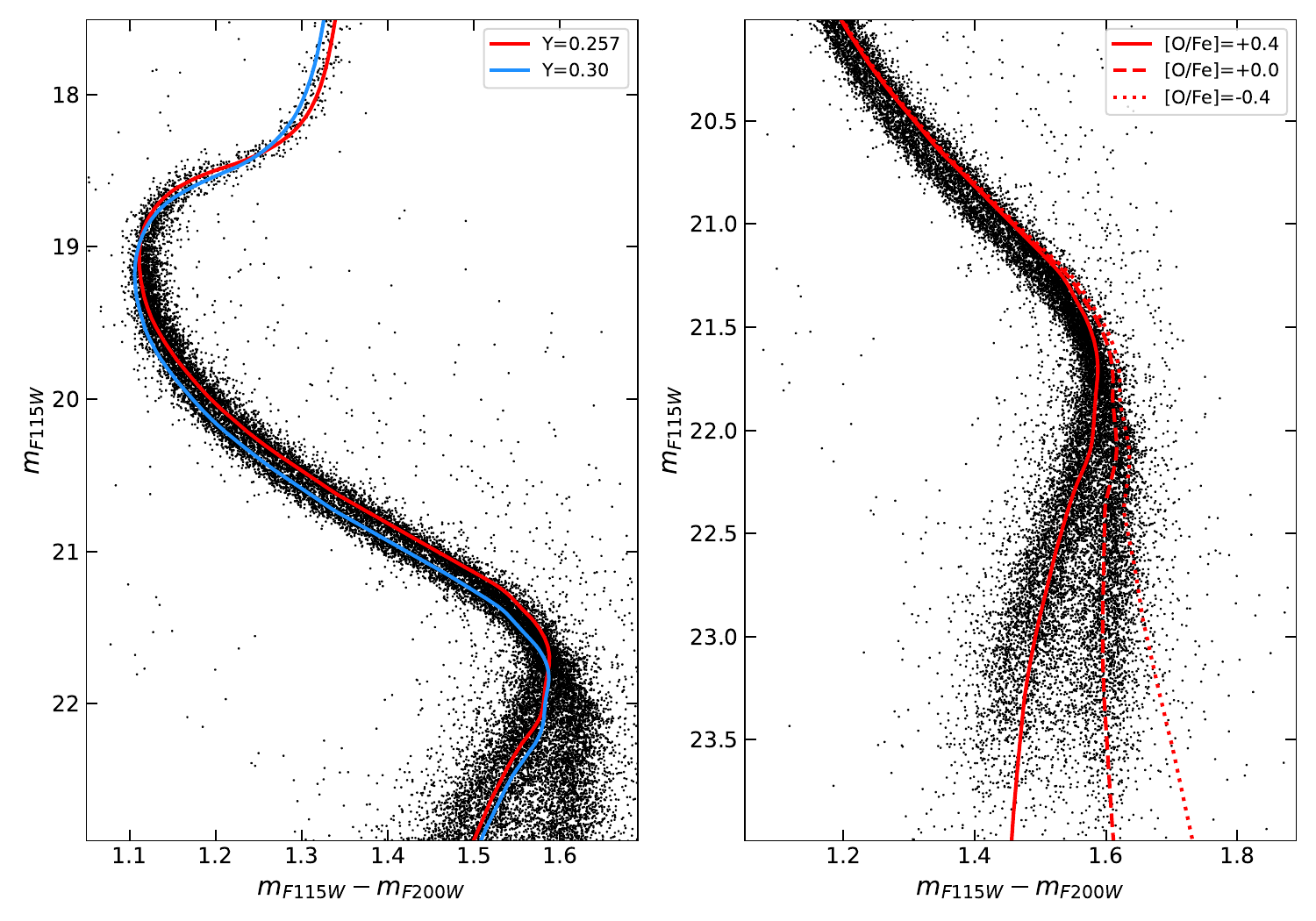}
\caption{{\it Left-hand panel:} JWST CMD of NGC 6440 including only bona-fide member stars. The red solid curve is a 13 Gyr BaSTI isochrone reproducing a $[\alpha/Fe]=+0.4$ stellar population with a metallicity $[Fe/H]=-0.6$ and a He fraction $Y=0.257$, representative of the FP. The blue curve is the same model but with a He fraction $Y=0.30$. {\it Right-hand panel:} same as in the left-hand panel but zoomed on the lower main-sequence. The solid curve is the same as in the left-hand panel and reproduces the FP ($O/Fe=+0.4$), while the dashed and dotted curves are representative of two different SPs with a solar-scaled ($O/Fe=+0.0$)  and ($O/Fe=-0.4$) oxygen abundance, respectively.}  
\label{fig5}
\end{figure*}

To quantify the degree of light-element abundance variations, we compared the observed evolutionary sequences with a set of isochrones. As reference model, we extracted a BaSTI-IAC isochrone \citep{hidalgo2018,pietrinferni21} with an  age of 13 Gyr, a distance modulus $(m-M)_0 = 14.65$, an extinction $E(B-V)=1.26$ \citep{pallanca19,pallanca21}, a metallicity $[Fe/H]=-0.6$ \citep{origlia08,munoz17,crociati23}\footnote{The adopted metallicity value is 0.05-0.1 dex smaller, than that derived through spectroscopy, but it better reproduces the observed extension and shape of the main-sequence. It is worth stressing, however, that all the following results remain unchanged if a metallicity 0.05-0.1 dex larger is adopted.} , a helium content $Y=0.257$, and an $\alpha$-enhanced chemical mixture with $[\alpha /Fe]=0.4$, as commonly observed in bulge GCs. Temperature dependent extinction coefficients were calculated following \citet{cardelli89} and assuming $R_V=2.7$ \citep{pallanca21}\footnote{Please note that a shift by -0.04 mag in the F115W was necessary to simultaneously match the evolutionary sequences in all the available filter combinations. Such an offset could be due to uncertainties in the aperture correction, zero points and encircled energy fractions (see discussion in Section~4 of \citealt{nardiello22}).}. This isochrone describes the CMD location of the FP and is depicted as a solid red curve in Figure~\ref{fig5}. It is evident that the model effectively reproduces the evolutionary sequence of the cluster, particularly the red side of the bi-modal upper main-sequence and the blue side of the lower main-sequence {(see also the red solid curve in the right-hand panel of Figure~\ref{fig5})}, as expected for an FP with standard He mass fraction.  We then extracted a second isochrone with the same properties as the first one but with a higher helium content of $Y=0.30$. This model (blue line in Figure~\ref{fig5})  aligns well with the observed evolutionary sequence and, specifically, adequately fits the blue side of the upper main-sequence. The color difference between the two isochrones progressively decreases for $m_{F115W}<20$ and reaches a minimum around the turn-off point, thus naturally explaining the non-detection of the bi-modality in the brighter main-sequence portion. Notably, the two models are also capable of explaining the observed width of the red-giant branch, which is significantly larger than expected from the photometric errors. 
In conclusion, the good match between the observed sequences and the models supports the conclusion that the upper bi-modality is indeed due to helium variations, suggesting an abundance spread   $\Delta Y \approx 0.04$. On the other hand, the helium variation have a negligible impact below the main-sequence knee and therefore cannot fully account for the lower bi-modality.

\begin{figure}[h] 
\centering
\includegraphics[scale=0.43]{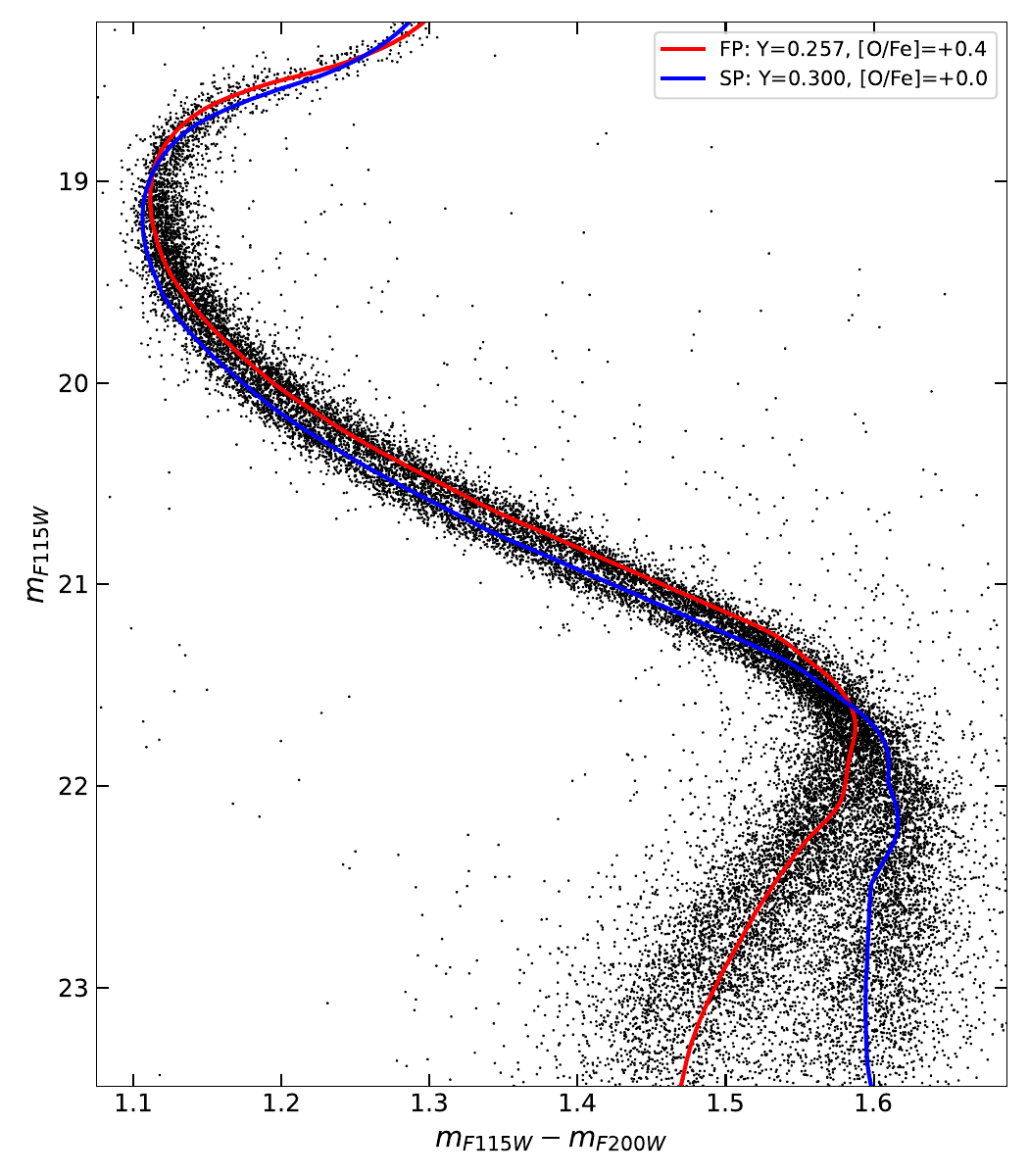}
\caption{JWST CMD of NGC 6440 including only bona-fide member stars. including only bona-fide member stars. The red curve is a 13 Gyr BaSTI isochrone reproducing a $[\alpha/Fe]=+0.4$ stellar population with a metallicity $[Fe/H]=-0.6$ and a He fraction $Y=0.257$, representative of the FP. The blue curve is an isochrone computed at the same age and metallicity as the previous one, but with a He fraction $Y=0.3$ and a solar-scaled ($O/Fe=+0.0$) chemical mixture, thus representative of the SP. }  
\label{fig6}
\end{figure}

To reproduce the bi-modality observed in the lower main-sequence, we computed a set of new isochrones with different choices for the metal distribution, to account for the typical observed anti-correlations among carbon, nitrogen and oxygen in the SPs. The calculations of the model atmospheres and fluxes followed the procedure described in \citet{hidalgo2018} using the latest version of the ATLAS9 code. Specifically, we computed spectra and magnitudes for a mildly O-depleted mixture with [O/Fe]=0.0, [C/Fe]$=-0.3$, and [N/Fe]$= +0.7$, and a highly O-depleted mixture with [O/Fe]$=-0.4$, [C/Fe]$=-0.6$, and [N/Fe]$=+1.4$ (the C + N + O abundance is kept constant). 
The FP isochrone and these two new SP isochrones with varying oxygen abundances are shown in the right-hand panel of Figure~\ref{fig5}. As expected, oxygen variations significantly alter the shape of the isochrones exclusively below the main-sequence knee. In particular, it is evident that the isochrone with [O/Fe]$=+0.0$ effectively matches the red branch of the bi-modal main-sequence, while the colours for the  [O/Fe]$=-0.4$ isochrone are too red. These findings suggest the the bi-modality below the knee is indeed due to the presence of an O-depleted population with $\Delta [O/Fe] \sim -0.4$ compared to FP stars. Finally, in Figure~\ref{fig6} we show the reference FP isochrone alongside an SP isochrone computed assuming a helium rich $Y=0.30$ and mildly O-depleted  $[O/Fe]=0.0$  chemical mixture. It is clear that these two models provide a nice fit to the whole cluster evolutionary sequence, successfully reproducing both the upper and lower bi-modalities.

It is however important to highlight that, as discussed by \citet{vandenberg23}, the chemical anti-correlations typical of MPs lead to effective temperature offsets among isochrones for very low-mass stars located below the main-sequence knee, particularly in the high metallicity regime of NGC 6440 stars. This is due to the effect of an O-depleted metal mixture on the boundary conditions for very low-mass stellar models.
These effective temperature offsets are not included in our SP isochrones due to the lack of appropriate boundary conditions with the adopted chemical mixtures in the BaSTI calculations of very low-mass stars. According to Figure~3 in \citet{vandenberg23}, the O-depletion results in a higher effective temperature of very low-mass stars, and therefore  neglecting this temperature effect implies that the value of $\Delta [O/Fe] \sim -0.4$ estimated here could be slightly underestimated.

\section{Summary and conclusions}
This Letter presents the first results of JWST observations
of the massive, dense, highly obscured, and highly field-contaminated bulge GC NGC~6440. We used ultra-deep observations acquired with the NIRCam in the F115W and F200W filters to construct a CMD extending down to $\sim5$ magnitudes below the main-sequence turn-off.  By exploiting the synergy with archival HST/WFC3 data, we carefully corrected the CMD for the differential extinction affecting the field of view and disentangled the cluster population from field interlopers through proper motion analysis. This allowed us to obtain the first photometric evidence of MPs in this observationally challenging cluster.

The presence of MPs is observed along the main-sequence, which clearly exhibits two distinct and unambiguous bi-modal patterns. One is observed at magnitudes brighter than the main-sequence knee ($m_{F115W}<22$) , and the other at magnitudes fainter than it. A similar behavior has also been observed using optical and near-IR observations in NGC~2808, a massive GC with extreme MP patterns \citep[e.g.,][]{piotto07,milone12}.

Above the knee, the main-sequence splits into two branches: a red one containing approximately 60\% of the stars and a blue one containing  the remaining 40\%. Through comparison with a suitable set of isochrones, we demonstrate that this bi-modality is most likely the result of the coexistence of a helium-standard ($Y=0.26$) and a helium-enriched stellar population with a difference $\Delta Y\approx0.04$. Although further dedicated modeling will be necessary to accurately quantify the helium enrichment history in this cluster, main-sequence splits attributed to such phenomenon have been observed in other clusters, s, such as $\omega$ Centauri, NGC~2808, and NGC~6441 \citep{piotto07,king12,bellini13}. Indeed,  the existence of heavily helium-enriched stellar populations appears to be a common feature in massive GCs, with the detected helium spreads showing an increasing trend for increasing cluster mass. The estimated helium enrichment is consistent with expectations for a massive cluster like NGC~6440 \citep{Milone2018}, and could also explain its horizontal branch morphology \citep{mauro12}.

Below the knee, the main-sequence splits again into at least two branches. A similar feature has been already observed among  low-mass stars (M-dwarfs) in GCs showing MPs and is likely due to differences in water (i.e., oxygen) abundance \citep[see, e.g.,][and references therein]{milone19}.  In the case of NGC~6440, the bi-modality consists of a blue branch, likely populated by FP (O-rich) stars, and a red one, likely populated by SP (O-depleted) stars, with the relative fraction of the two population stars being approximately consistent with that estimated above the main sequence knee. The blue sequence appears to be further structured into two sub-sequences. Although this feature currently lacks a statistical confirmation, it could suggest the presence of a more complex MP pattern in this cluster which is worth further investigations in the future. By comparing the observed sequences with a suitable set of isochrones adopting different chemical mixtures typical of MPs, we confirm that the observed bi-modality is due to oxygen variations and we suggest an abundance spread of approximately $\Delta [O/Fe]\sim-0.4$. 

Evidence of MPs below the knee was previously obtained with JWST in the case of M~92 and NGC~104 \citep{nardiello22,milone2023}. However, a combination of JWST data (acquired with photometric filters different from those used here) and HST optical/near-UV observations has been necessary for the MP characterization in those clusters. Despite the similar metallicity between NGC~6440 and NGC~104, the present work clearly shows, instead, that the JWST F115W-F200W filter combination is able to unveil the presence of MPs both above and below the main-sequence knee, producing a  striking bi-modal colour distribution, with no need of additional HST photometry. 
This highlights the stand-alone effectiveness of these two JWST filters in discriminating MPs. The results presented in this Letter demonstrate how the superb capabilities of the JWST can be exploited to systematically map the MP phenomenon in observationally challenging clusters, such as those located in the Galactic bulge. Such mapping was not feasible through spectroscopy of large star samples or near-UV (or even optical) observations  and holds the potential to investigate the MP phenomeon in a different chemical and dynamical environment mostly populated by in-situ formed GCs.

\begin{acknowledgements}
      M.C., C.P., E.D., S.L., F.R.F and B.L. acknowledge financial support from the project Light-on-Dark granted by MIUR through PRIN2017-2017K7REXT. M.S. acknowledges support from The Science and Technology Facilities Council Consolidated Grant ST/V00087X/1. C.H. is supported by NSERC Discovery grant RGPIN-2023-04264.This research is part of the project Cosmic-Lab (Globular Clusters as Cosmic Laboratories) at the Physics and Astronomy Department A. Righi of the Bologna University (http://www.cosmic-lab.eu/Cosmic-Lab/Home.html).
\end{acknowledgements}

%
\bibliographystyle{aa} 
\bibliography{biblio_n6440} 

\begin{thebibliography}{49}
\expandafter\ifx\csname natexlab\endcsname\relax\def\natexlab#1{#1}\fi

\bibitem[{{Bastian} \& {Lardo}(2018)}]{Bastian2018}
{Bastian}, N. \& {Lardo}, C. 2018, \araa, 56, 83

\bibitem[{{Bellini} {et~al.}(2013){Bellini}, {Piotto}, {Milone}, {King},
  {Renzini}, {Cassisi}, {Anderson}, {Bedin}, {Nardiello}, {Pietrinferni}, \&
  {Sarajedini}}]{bellini13}
{Bellini}, A., {Piotto}, G., {Milone}, A.~P., {et~al.} 2013, \apj, 765, 32

\bibitem[{{Bellini} {et~al.}(2015){Bellini}, {Renzini}, {Anderson}, {Bedin},
  {Piotto}, {Soto}, {Brown}, {Milone}, {Sohn}, \& {Sweigart}}]{Bellini2015}
{Bellini}, A., {Renzini}, A., {Anderson}, J., {et~al.} 2015, \apj, 805, 178

\bibitem[{{Cadelano} {et~al.}(2017){Cadelano}, {Dalessandro}, {Ferraro},
  {Miocchi}, {Lanzoni}, {Pallanca}, \& {Massari}}]{cadelano17}
{Cadelano}, M., {Dalessandro}, E., {Ferraro}, F.~R., {et~al.} 2017, \apj, 836,
  170

\bibitem[{{Cadelano} {et~al.}(2022){Cadelano}, {Dalessandro}, {Salaris},
  {Bastian}, {Mucciarelli}, {Saracino}, {Martocchia}, \&
  {Cabrera-Ziri}}]{cadelano2022}
{Cadelano}, M., {Dalessandro}, E., {Salaris}, M., {et~al.} 2022, \apjl, 924, L2

\bibitem[{{Cadelano} {et~al.}(2020){Cadelano}, {Saracino}, {Dalessandro},
  {Ferraro}, {Lanzoni}, {Massari}, {Pallanca}, \& {Salaris}}]{cadelano20_n6256}
{Cadelano}, M., {Saracino}, S., {Dalessandro}, E., {et~al.} 2020, \apj, 895, 54

\bibitem[{{Cardelli} {et~al.}(1989){Cardelli}, {Clayton}, \&
  {Mathis}}]{cardelli89}
{Cardelli}, J.~A., {Clayton}, G.~C., \& {Mathis}, J.~S. 1989, \apj, 345, 245

\bibitem[{{Cassisi} \& {Salaris}(2020)}]{cassisi20}
{Cassisi}, S. \& {Salaris}, M. 2020, \aapr, 28, 5

\bibitem[{{Crociati} {et~al.}(2023){Crociati}, {Valenti}, {Ferraro},
  {Pallanca}, {Lanzoni}, {Cadelano}, {Fanelli}, {Origlia}, {Leanza},
  {Dalessandro}, {Mucciarelli}, \& {Rich}}]{crociati23}
{Crociati}, C., {Valenti}, E., {Ferraro}, F.~R., {et~al.} 2023, \apj, 951, 17

\bibitem[{{Dalessandro} {et~al.}(2019){Dalessandro}, {Cadelano}, {Vesperini},
  {Martocchia}, {Ferraro}, {Lanzoni}, {Bastian}, {Hong}, \&
  {Sanna}}]{dalessandro2019}
{Dalessandro}, E., {Cadelano}, M., {Vesperini}, E., {et~al.} 2019, \apjl, 884,
  L24

\bibitem[{{Dalessandro} {et~al.}(2018{\natexlab{a}}){Dalessandro}, {Cadelano},
  {Vesperini}, {Salaris}, {Ferraro}, {Lanzoni}, {Raso}, {Hong}, {Webb}, \&
  {Zocchi}}]{dalessandro18_m80}
{Dalessandro}, E., {Cadelano}, M., {Vesperini}, E., {et~al.}
  2018{\natexlab{a}}, \apj, 859, 15

\bibitem[{{Dalessandro} {et~al.}(2016){Dalessandro}, {Lapenna}, {Mucciarelli},
  {Origlia}, {Ferraro}, \& {Lanzoni}}]{Dalessandro2016}
{Dalessandro}, E., {Lapenna}, E., {Mucciarelli}, A., {et~al.} 2016, \apj, 829,
  77

\bibitem[{{Dalessandro} {et~al.}(2018{\natexlab{b}}){Dalessandro}, {Lardo},
  {Cadelano}, {Saracino}, {Bastian}, {Mucciarelli}, {Salaris}, {Stetson}, \&
  {Pancino}}]{dalessandro2018}
{Dalessandro}, E., {Lardo}, C., {Cadelano}, M., {et~al.} 2018{\natexlab{b}},
  \aap, 618, A131

\bibitem[{{Ferraro} {et~al.}(2023){Ferraro}, {Lanzoni}, {Vesperini},
  {Cadelano}, {Deras}, \& {Pallanca}}]{ferraro23}
{Ferraro}, F.~R., {Lanzoni}, B., {Vesperini}, E., {et~al.} 2023, \apj, 950, 145

\bibitem[{{Gratton} {et~al.}(2019){Gratton}, {Bragaglia}, {Carretta},
  {D'Orazi}, {Lucatello}, \& {Sollima}}]{gratton19}
{Gratton}, R., {Bragaglia}, A., {Carretta}, E., {et~al.} 2019, \aapr, 27, 8

\bibitem[{Hartigan \& Hartigan(1985)}]{dip85}
Hartigan, J.~A. \& Hartigan, P.~M. 1985, The Annals of Statistics, 13, 70

\bibitem[{{Hidalgo} {et~al.}(2018){Hidalgo}, {Pietrinferni}, {Cassisi},
  {Salaris}, {Mucciarelli}, {Savino}, {Aparicio}, {Silva Aguirre}, \&
  {Verma}}]{hidalgo2018}
{Hidalgo}, S.~L., {Pietrinferni}, A., {Cassisi}, S., {et~al.} 2018, \apj, 856,
  125

\bibitem[{{King} {et~al.}(2012){King}, {Bedin}, {Cassisi}, {Milone}, {Bellini},
  {Piotto}, {Anderson}, {Pietrinferni}, \& {Cordier}}]{king12}
{King}, I.~R., {Bedin}, L.~R., {Cassisi}, S., {et~al.} 2012, \aj, 144, 5

\bibitem[{{Larsen} {et~al.}(2014){Larsen}, {Brodie}, {Grundahl}, \&
  {Strader}}]{larsen14}
{Larsen}, S.~S., {Brodie}, J.~P., {Grundahl}, F., \& {Strader}, J. 2014, \apj,
  797, 15

\bibitem[{{Leanza} {et~al.}(2023){Leanza}, {Pallanca}, {Ferraro}, {Lanzoni},
  {Dalessandro}, {Cadelano}, {Vesperini}, {Origlia}, {Mucciarelli}, \&
  {Valenti}}]{leanza23}
{Leanza}, S., {Pallanca}, C., {Ferraro}, F.~R., {et~al.} 2023, \apj, 944, 162

\bibitem[{{Martocchia} {et~al.}(2018){Martocchia}, {Niederhofer},
  {Dalessandro}, {Bastian}, {Kacharov}, {Usher}, {Cabrera-Ziri}, {Lardo},
  {Cassisi}, {Geisler}, {Hilker}, {Hollyhead}, {Kozhurina-Platais}, {Larsen},
  {Mackey}, {Mucciarelli}, {Platais}, \& {Salaris}}]{Martocchia2018}
{Martocchia}, S., {Niederhofer}, F., {Dalessandro}, E., {et~al.} 2018, \mnras,
  477, 4696

\bibitem[{{Massari} {et~al.}(2019){Massari}, {Koppelman}, \&
  {Helmi}}]{massari19}
{Massari}, D., {Koppelman}, H.~H., \& {Helmi}, A. 2019, \aap, 630, L4

\bibitem[{{Mauro} {et~al.}(2012){Mauro}, {Moni Bidin}, {Cohen}, {Geisler},
  {Minniti}, {Catelan}, {Chen{\'e}}, \& {Villanova}}]{mauro12}
{Mauro}, F., {Moni Bidin}, C., {Cohen}, R., {et~al.} 2012, \apjl, 761, L29

\bibitem[{{Milone} {et~al.}(2019{\natexlab{a}}){Milone}, {Marino}, {Bedin},
  {Anderson}, {Apai}, {Bellini}, {Dieball}, {Salaris}, {Libralato},
  {Nardiello}, {Bergeron}, {Burgasser}, {Rees}, {Rich}, \&
  {Richer}}]{milone2019}
{Milone}, A.~P., {Marino}, A.~F., {Bedin}, L.~R., {et~al.} 2019{\natexlab{a}},
  \mnras, 484, 4046

\bibitem[{{Milone} {et~al.}(2019{\natexlab{b}}){Milone}, {Marino}, {Bedin},
  {Anderson}, {Apai}, {Bellini}, {Dieball}, {Salaris}, {Libralato},
  {Nardiello}, {Bergeron}, {Burgasser}, {Rees}, {Rich}, \& {Richer}}]{milone19}
{Milone}, A.~P., {Marino}, A.~F., {Bedin}, L.~R., {et~al.} 2019{\natexlab{b}},
  \mnras, 484, 4046

\bibitem[{{Milone} {et~al.}(2012){Milone}, {Marino}, {Cassisi}, {Piotto},
  {Bedin}, {Anderson}, {Allard}, {Aparicio}, {Bellini}, {Buonanno}, {Monelli},
  \& {Pietrinferni}}]{milone12}
{Milone}, A.~P., {Marino}, A.~F., {Cassisi}, S., {et~al.} 2012, \apjl, 754, L34

\bibitem[{{Milone} {et~al.}(2023){Milone}, {Marino}, {Dotter}, {Ziliotto},
  {Dondoglio}, {Cordoni}, {Jang}, {Lagioia}, {Legnardi}, {Mohandasan}, {Tailo},
  {Yong}, {Baimukhametova}, \& {Carlos}}]{milone2023}
{Milone}, A.~P., {Marino}, A.~F., {Dotter}, A., {et~al.} 2023, \mnras, 522,
  2429

\bibitem[{{Milone} {et~al.}(2018){Milone}, {Marino}, {Renzini}, {D'Antona},
  {Anderson}, {Barbuy}, {Bedin}, {Bellini}, {Brown}, {Cassisi}, {Cordoni},
  {Lagioia}, {Nardiello}, {Ortolani}, {Piotto}, {Sarajedini}, {Tailo}, {van der
  Marel}, \& {Vesperini}}]{Milone2018}
{Milone}, A.~P., {Marino}, A.~F., {Renzini}, A., {et~al.} 2018, \mnras, 481,
  5098

\bibitem[{{Milone} {et~al.}(2017){Milone}, {Piotto}, {Renzini}, {Marino},
  {Bedin}, {Vesperini}, {D'Antona}, {Nardiello}, {Anderson}, {King}, {Yong},
  {Bellini}, {Aparicio}, {Barbuy}, {Brown}, {Cassisi}, {Ortolani}, {Salaris},
  {Sarajedini}, \& {van der Marel}}]{Milone2017}
{Milone}, A.~P., {Piotto}, G., {Renzini}, A., {et~al.} 2017, \mnras, 464, 3636

\bibitem[{{Mu{\~n}oz} {et~al.}(2017){Mu{\~n}oz}, {Villanova}, {Geisler},
  {Saviane}, {Dias}, {Cohen}, \& {Mauro}}]{munoz17}
{Mu{\~n}oz}, C., {Villanova}, S., {Geisler}, D., {et~al.} 2017, \aap, 605, A12

\bibitem[{{Mucciarelli} {et~al.}(2008){Mucciarelli}, {Carretta}, {Origlia}, \&
  {Ferraro}}]{mucciarelli08}
{Mucciarelli}, A., {Carretta}, E., {Origlia}, L., \& {Ferraro}, F.~R. 2008,
  \aj, 136, 375

\bibitem[{{Nardiello} {et~al.}(2022){Nardiello}, {Bedin}, {Burgasser},
  {Salaris}, {Cassisi}, {Griggio}, \& {Scalco}}]{nardiello22}
{Nardiello}, D., {Bedin}, L.~R., {Burgasser}, A., {et~al.} 2022, \mnras, 517,
  484

\bibitem[{{O'Donnell}(1994)}]{O'Donnell1994}
{O'Donnell}, J.~E. 1994, \apj, 422, 158

\bibitem[{{Onorato} {et~al.}(2023){Onorato}, {Cadelano}, {Dalessandro},
  {Vesperini}, {Lanzoni}, \& {Mucciarelli}}]{onorato23}
{Onorato}, S., {Cadelano}, M., {Dalessandro}, E., {et~al.} 2023, \aap, 677, A8

\bibitem[{{Origlia} {et~al.}(2008){Origlia}, {Valenti}, \& {Rich}}]{origlia08}
{Origlia}, L., {Valenti}, E., \& {Rich}, R.~M. 2008, \mnras, 388, 1419

\bibitem[{{Pallanca} {et~al.}(2019){Pallanca}, {Ferraro}, {Lanzoni},
  {Saracino}, {Raso}, \& {Focardi}}]{pallanca19}
{Pallanca}, C., {Ferraro}, F.~R., {Lanzoni}, B., {et~al.} 2019, \apj, 882, 159

\bibitem[{{Pallanca} {et~al.}(2021){Pallanca}, {Lanzoni}, {Ferraro},
  {Casagrande}, {Saracino}, {Purohith Bhaskar Bhat}, {Leanza}, {Dalessandro},
  \& {Vesperini}}]{pallanca21}
{Pallanca}, C., {Lanzoni}, B., {Ferraro}, F.~R., {et~al.} 2021, \apj, 913, 137

\bibitem[{{Pietrinferni} {et~al.}(2021){Pietrinferni}, {Hidalgo}, {Cassisi},
  {Salaris}, {Savino}, {Mucciarelli}, {Verma}, {Silva Aguirre}, {Aparicio}, \&
  {Ferguson}}]{pietrinferni21}
{Pietrinferni}, A., {Hidalgo}, S., {Cassisi}, S., {et~al.} 2021, \apj, 908, 102

\bibitem[{{Piotto} {et~al.}(2007){Piotto}, {Bedin}, {Anderson}, {King},
  {Cassisi}, {Milone}, {Villanova}, {Pietrinferni}, \& {Renzini}}]{piotto07}
{Piotto}, G., {Bedin}, L.~R., {Anderson}, J., {et~al.} 2007, \apjl, 661, L53

\bibitem[{{Piotto} {et~al.}(2015{\natexlab{a}}){Piotto}, {Milone}, {Bedin},
  {Anderson}, {King}, {Marino}, {Nardiello}, {Aparicio}, {Barbuy}, {Bellini},
  {Brown}, {Cassisi}, {Cool}, {Cunial}, {Dalessandro}, {D'Antona}, {Ferraro},
  {Hidalgo}, {Lanzoni}, {Monelli}, {Ortolani}, {Renzini}, {Salaris},
  {Sarajedini}, {van der Marel}, {Vesperini}, \& {Zoccali}}]{piotto15}
{Piotto}, G., {Milone}, A.~P., {Bedin}, L.~R., {et~al.} 2015{\natexlab{a}},
  \aj, 149, 91

\bibitem[{{Piotto} {et~al.}(2015{\natexlab{b}}){Piotto}, {Milone}, {Bedin},
  {Anderson}, {King}, {Marino}, {Nardiello}, {Aparicio}, {Barbuy}, {Bellini},
  {Brown}, {Cassisi}, {Cool}, {Cunial}, {Dalessandro}, {D'Antona}, {Ferraro},
  {Hidalgo}, {Lanzoni}, {Monelli}, {Ortolani}, {Renzini}, {Salaris},
  {Sarajedini}, {van der Marel}, {Vesperini}, \& {Zoccali}}]{piotto2015}
{Piotto}, G., {Milone}, A.~P., {Bedin}, L.~R., {et~al.} 2015{\natexlab{b}},
  \aj, 149, 91

\bibitem[{{Renzini} {et~al.}(2022){Renzini}, {Marino}, \& {Milone}}]{renzini22}
{Renzini}, A., {Marino}, A.~F., \& {Milone}, A.~P. 2022, \mnras, 513, 2111

\bibitem[{{Salaris} {et~al.}(2019){Salaris}, {Cassisi}, {Mucciarelli}, \&
  {Nardiello}}]{salaris19}
{Salaris}, M., {Cassisi}, S., {Mucciarelli}, A., \& {Nardiello}, D. 2019, \aap,
  629, A40

\bibitem[{{Salaris} {et~al.}(2006){Salaris}, {Weiss}, {Ferguson}, \&
  {Fusilier}}]{salaris06}
{Salaris}, M., {Weiss}, A., {Ferguson}, J.~W., \& {Fusilier}, D.~J. 2006, \apj,
  645, 1131

\bibitem[{{Sbordone} {et~al.}(2011){Sbordone}, {Salaris}, {Weiss}, \&
  {Cassisi}}]{Sbordone2011}
{Sbordone}, L., {Salaris}, M., {Weiss}, A., \& {Cassisi}, S. 2011, \aap, 534,
  A9

\bibitem[{{Sills} {et~al.}(2019){Sills}, {Dalessandro}, {Cadelano},
  {Alfaro-Cuello}, \& {Kruijssen}}]{sills19}
{Sills}, A., {Dalessandro}, E., {Cadelano}, M., {Alfaro-Cuello}, M., \&
  {Kruijssen}, J.~M.~D. 2019, \mnras, 490, L67

\bibitem[{{Stetson}(1987)}]{Stetson1987}
{Stetson}, P.~B. 1987, \pasp, 99, 191

\bibitem[{{Stetson}(1994)}]{Stetson1994}
{Stetson}, P.~B. 1994, \pasp, 106, 250

\bibitem[{{VandenBerg}(2023)}]{vandenberg23}
{VandenBerg}, D.~A. 2023, \mnras, 518, 4517

\end{thebibliography}
%

\end{document}